\documentclass[aps,preprint,showpacs,a4]{revtex4-1}
\usepackage{graphicx}
\newcommand{\e}{\textnormal{e}}
\usepackage{verbatim,amsmath}

\begin{document}

\date{\today)}

\title[Quantum noise]{Reduction of quantum noise in optical interferometers using squeezed light}

\author{Michael Weyrauch}\email{michael.weyrauch@ptb.de}
\affiliation{Physikalisch-Technische Bundesanstalt, D-38116
Braunschweig, Germany}

\author{Volodymyr G. Voronov}\email{v\_voronov@univ.kiev.ua}
\affiliation{Faculty of Physics, Taras Shevchenko National
University of Kyiv, Kyiv, Ukraine}

\begin{abstract}
We study the photon counting noise in optical interferometers used
for gravitational wave detection. In order to reduce quantum noise a
squeezed vacuum state is injected into the usually unused input
port. Here, we specifically investigate the so called `dark port
case', when the beam splitter is oriented close to 90$^{\circ}$ to
the incoming laser beam, such that nearly all photons  go to one
output port of the interferometer, and only a small fraction of
photons is seen in the other port (`dark port'). For this case it
had been suggested that signal amplification is possible without
concurrent noise amplification [R.~Barak and Y.~Ben-Aryeh,
J.~Opt.~Soc.~Am.~B25(361)2008]. We show that by injection of a
squeezed vacuum state into the second input port, counting noise is
reduced for large values of the squeezing factor, however the signal
is not amplified. Signal strength only depends on the intensity of
the laser beam.

\end{abstract}

\maketitle

\section{Introduction}\label{sec:introduction}

For gravitational wave detection with optical interferometers
various sources of noise must be carefully controlled and, if
possible, minimized. In an effort to reduce quantum-mechanical
noise, Caves~\cite{caves:1} proposed the squeezed state technique:
into the normally unused port of the interferometer a squeezed
vacuum state is injected. Details of this technique are analyzed
e.g. in Refs.~\cite{yurke:1, assaf:1} and references therein.

The photon state in the interferometer after passing a beam splitter
is not a product state, but the states of the output ports are
entangled. In recent papers by Barak and Ben-Aryeh~\cite{barak:1}
and Voronov and Weyrauch~\cite{voronov:1}, the consequences of this
entanglement for the photon statistics of an optical interferometer
were studied.

In Ref.~\cite{barak:1} it was suggested that under certain
conditions, the gravitational wave signal may by amplified without a
corresponding increase in counting noise. In Ref.~\cite{voronov:1}
we disputed this surprising prediction, and showed that is was the
result of an inaccuracy in the calculations. Furthermore, we
calculated photon distributions in the output state for various
settings of a beam splitter with respect to weak and strong incoming
laser fields. We showed that a squeezed vacuum injected into the
other port cannot amplify the signal however my reduce counting
noise for large squeezing factors.

In a recent paper~\cite{ben_aryeh:1}, Ben-Aryeh specifically
reanalyzes the `dark port case', i.e. a configuration where the beam
splitter is oriented close to 90$^\circ$ to the incoming laser beam
with a squeezed vacuum entering the other port. He confirms our
findings in Ref.~\cite{voronov:1} and sharpens the physical
interpretation of the results obtained.

It is the purpose of the present paper to investigate the photon
statistics in the dark output port of the interferometer in more
detail, and present the calculations and results more succinctly
than in our previous paper~\cite{voronov:1}.

In section~\ref{sec:approximation} we develop formulas for the
calculation of the photon number distributions in the dark output
port of the interferometer, as well as their mean values and
variances. Numerical results and their physical interpretation will
be discussed in section 3. A brief summary concludes the paper.

\section{Photon statistics in dark output port of a beam splitter}
\label{sec:approximation}

The photon field operators $\hat{a}_i$ and $\hat{a}_i^{\dagger}$ of
the input ports and the photon field operators  $\hat{b}_i$ and
$\hat{b}_j^\dagger$ of the output ports of a beam splitter are
related through the beam splitter transformation~\cite{campos:1}
\begin{equation}
\left(
  \begin{array}{c}
    \hat{a}_1 \\
    \hat{a}_2 \\
  \end{array}
\right)=
\left(
  \begin{array}{cc}
    \cos\gamma & \sin\gamma \\
    -\sin\gamma & \cos\gamma \\
  \end{array}
\right)
\left(
  \begin{array}{c}
    \hat{b}_1 \\
    \hat{b}_2 \\
  \end{array}
\right).
\end{equation}
Both sets of operators fulfill boson commutation relations
$[\hat{a}_i,\hat{a}_j^\dagger]=\delta_{i,j}$ and
$[\hat{b}_i,\hat{b}_j^\dagger]=\delta_{i,j}$, respectively. The
parameter $\gamma$ parameterizes the splitting ratio of the beam
splitter with respect to the incoming laser beam.

The incoming laser beam in port 1 is a coherent state, and a
squeezed vacuum state is injected in port 2:
\begin{equation}
|\psi_{\rm in}(\alpha,\zeta)\rangle = \hat{S}_2(\zeta)\hat{D}_1(\alpha)|0,0\rangle
\end{equation}
with
\begin{equation}
\hat{D}_1(\alpha)=\exp\left(\alpha \hat{a}_1^\dagger -\alpha^*\hat{a}_1\right),\;\;\;
\hat{S}_2(\zeta)=\exp\left(\frac{\zeta^*}{2} \hat{a}_2^{ 2} -\frac{\zeta}{2} \hat{a}_{2}^{\dagger2}\right).
\end{equation}
The coherence parameter $\alpha$ and the squeezing parameter $\zeta$
are complex numbers.

The beam splitter transformation allows to write the $\hat{a}$
operators in terms of the $\hat{b}$ operators, and one may write the
photon state after passing the beam splitter as
\begin{equation}\label{eq:output}
|\psi_{\rm out}(\alpha,\zeta,\gamma)\rangle = \exp(|\zeta|\hat{A}) \hat{D}_1(\alpha\cos\gamma) \hat{D}_2(\alpha\sin\gamma) |0,0\rangle
\end{equation}
with
\begin{equation}\label{eq:A}
\hat{A}=\hat{s}_1 \sin^2\gamma+\hat{s}_2\cos^2\gamma+\hat{s}_{12}\sin\gamma\cos\gamma
\end{equation}
and
\begin{equation}\label{eq:oper1}
\hat{s}_i=\frac{1}{2|\zeta|} (\zeta^* \hat{b}_i^{2}-\zeta \hat{b}_i^{\dagger2})\;\;\;\;\;\;\;
\hat{s}_{12}=\frac{1}{|\zeta|}(\zeta \hat{b}_1^\dagger \hat{b}_2^\dagger
     -\zeta^* \hat{b}_1 \hat{b}_2 ).
\end{equation}

From the expression for $\hat{A}$ we see that both output states are
entangled by the operator $\hat{s}_{12}$. This fact significantly
complicates evaluation of the photon statistics of the output state.
However, it is possible to use Lie algebraic disentangling
techniques in order to rewrite the output state in a way which
enables the determination of photon distributions (for details we
refer the reader to Ref.~\cite{voronov:1}).

After disentangling it is possible to write Eq.~(\ref{eq:output}) as
follows~\cite{wei:1, scholz:2}
\begin{equation}\label{eq:output2}
|\psi_{\rm out}\rangle = \exp(\sigma_{T} \hat{t}_{12})\exp(\sigma_{S} \hat{s}_{12})\exp(\sigma_1\hat{s}_1)\exp(\sigma_2\hat{s}_2)
\hat{D}_1(\alpha \cos\gamma) \hat{D}_2(\alpha\sin\gamma)
|0,0\rangle
\end{equation}
with $\hat{t}_{12}= \hat{b}_1 \hat{b}_2^\dagger- \hat{b}_1^\dagger
\hat{b}_2$. The output state is now expressed in terms of two
squeezed coherent states entangled via the operators
$\exp(\sigma_{T} \hat{t}_{12})$ and $\exp(\sigma_{S} \hat{s}_{12})$.
The coefficients $\sigma_{T}$, $\sigma_{S}$, $\sigma_1$, $\sigma_2$
are real functions of the input parameters $r=|\zeta|$ and $\gamma$.
A simple method for the numerical determination of these parameters
is described in Appendix A of Ref.~\cite{voronov:1}.

The dark port case corresponds $\gamma=\pi/2-\delta$, where $\delta$
is a small phase shift (we assume it is real), for which one finds
to first order in $\delta$ (see Appendix A in Ref.~\cite{voronov:1})
\begin{equation}\label{eq:dis_dark}
\sigma_1=r, \ \ \sigma_2=0,\ \
\sigma_S=-\delta\sinh r, \ \ \text{and} \ \  \sigma_T=\delta(1-\cosh r).
\end{equation}
We furthermore assume a very strong coherent state incoming in port
1, such that the $\hat{b}_2$ and $\hat{b}_2^\dagger$ operators can
be replaced in the entanglement factors in Eq.~(\ref{eq:output2}) by
their expectation values $\alpha$ and $\alpha^*$, respectively. The
output state can then be written as
\begin{equation}\label{eq:app_out}
|\psi_{\rm out}\rangle=\hat{D}_1(-\alpha\delta(1-\cosh r))\hat{D}_1(-\delta\alpha^* \e^{i\theta}\sinh r)
\hat{S}_1(\zeta)\hat{D}_1(-\alpha \delta)\hat{D}_2(\alpha)|0,0\rangle
\end{equation}
with $\zeta=r \e^{i\theta}$.  The operators with index 1 may be
combined using the relations
\begin{eqnarray}
\hat{D}(\alpha_2)\hat{D}(\alpha_1)&=&\hat{D}(\alpha_1+\alpha_2) \exp \left[\frac{1}{2}(\alpha_2 \alpha_1^* - \alpha_2^*\alpha_1) \right],\nonumber\\
\hat{D}(\alpha)\hat{S}(\zeta)&=&\hat{S}(\zeta)\hat{D}(\alpha \cosh r + \alpha^* \e^{i \theta}\sinh r).
\end{eqnarray}
We finally obtain
\begin{equation}\label{eq:outpihalf}
|\psi_{\rm out}\rangle=\e^{i|\alpha|^2 \delta^2\Delta}\hat{S}_1(\zeta)\hat{D}_1(\tilde{\alpha})\hat{D}_2(\alpha)|0,0\rangle
\end{equation}
with $\alpha= |\alpha| \e^{i\phi}$ and
\begin{equation}\label{eq:alphatilde}
\begin{split}
&\tilde{\alpha}=-\alpha \delta \cosh r - \alpha^* \delta \e^{i \theta}\sinh r , \\
& \Delta=\frac{1}{2}\sin(\theta-2\phi)\sinh(2 r).
\end{split}
\end{equation}
As one can see from Eq.~(\ref{eq:outpihalf}), a strong coherent
state with coherence parameter $\alpha$ exits through port 2 of the
interferometer and a  weak squeezed coherent state with coherence
parameter $\tilde{\alpha}$ and squeezing parameter $\zeta$ exits
through port 1. Note that in Ref.~\cite{voronov:1} there are two
missprints: $\Delta$ must be defined without the term $\e^{2 i
\phi}$ and with opposite sign. Also note, that the coherence
parameter $\tilde{\alpha}$ depends on the squeezing parameter
$\zeta$, the coherence parameter $\alpha=|\alpha|\e^{i \phi}$ and
the phase shift $\delta$. The phase factor $\e^{i|\alpha|^2
\delta^2\Delta}$ is irrelevant for the determination of the photon
statistics.

In order to determine the photon statistics of the output state we
need to determine its number (Fock) representation. In terms of the
number representation of a squeezed coherent state~\cite{scully:1}
\begin{equation}\label{eq:focksqch}
\hat{S}(\zeta) \hat{D}(\tilde{\alpha})|0\rangle=
\sum_{n=0}^\infty \frac{1}{\sqrt{n!}}f_{n}(\zeta,\tilde{\alpha})  |n\rangle
\end{equation}
with
\begin{equation}
f_{n}(\zeta,\tilde{\alpha})=\frac{(\e^{i\theta}\tanh r)^{n/2}}{2^{n/2} (\cosh r)^{1/2}}
\exp\left(-\frac{1}{2}(|\tilde{\alpha}|^2-\e^{-i\theta}\tilde{\alpha}^2 \tanh r)\right) H_n\left(\frac{\tilde{\alpha}\e^{-i\theta/2}}
     {\sqrt{2\cosh r \sinh r}}\right),
\end{equation}
and $H_n$ the Hermite polynomials, one immediately  obtains the
distribution function
\begin{equation}\label{eq:darkdist}
P_{n_1}=\frac{1}{n_1!}\left|f_{n_1}(\zeta,\tilde{\alpha})\right|^2.
\end{equation}
The mean and the variance of this distribution may be obtained
analytically~\cite{scully:1}
\begin{equation}\label{eq:meanvari}
\begin{split}
\langle n_1 \rangle  &= |\alpha|^2 \delta^2+ \sinh^2 r, \\
 (\Delta n_1)^2  &= |\alpha|^2 \delta^2 (\cosh(2r) -\cos(\theta-2\phi)\sinh(2r)) +2 \sinh^2 r \cosh^2 r.
\end{split}
\end{equation}
Note, that the mean does not depend on the phases $\theta$ and
$\phi$, but the variance does. The results Eq.~(\ref{eq:meanvari})
on first sight appear different from those obtained in
Ref.~\cite{voronov:1}, however, it is possible to show that they are
equivalent. The form presented here is, however, much more
transparent.

Finally, we would like to remark, that the mean and variance
Eq.~(\ref{eq:meanvari}) may alternatively be calculated by a method
used by Caves~\cite{caves:1}. He expresses the output observables,
which are described by $\hat{b}$ operators, in terms of the input
operators $\hat{a}$ using the beam splitter transformation. In this
way one obtains the same results Eq.~(\ref{eq:meanvari}) very
efficiently, however, the full distribution function is not easily
obtained.

\section{Numerical results and physical interpretation}\label{se:numres}

We consider squeezing factors up to
$r=1.5$ in our numerical work, since at the present time the largest
squeezing factor experimentally realized is about $r=1.3$
corresponding to a maximum squeezing of about -11.5
dB~\cite{mehmet:1}.

From the formulas~(\ref{eq:darkdist}) and (\ref{eq:meanvari}) we see
that the mean of the photon distribution does not depend on the
phases $\theta$ and $\phi$, but the variance depends on the phase
relation $\theta-2\phi$. Obviously, choosing $\theta-2\phi=0$ is
optimal in the sense of minimizing  the counting noise.

Substituting  $\theta-2\phi=0$ From Eqs.~(\ref{eq:meanvari}) one
finds
\begin{equation}\label{eq:meanvari_case}
\begin{split}
& \langle n_1 \rangle =\delta^2|\alpha|^2 + \sinh^2 r, \\
& (\Delta n_1)^2
= \delta^2|\alpha|^2 \e^{-2r} +2 \sinh^2 r \cosh^2r.
\end{split}
\end{equation}
For strong coherent input state (large $|\alpha|$) and within the
squeezing factor ranges experimentally accessible ($r$ up to about
$1.3$) the second terms in both expressions can be neglected.
Consequently, amplification of the signal in output port 1 by
squeezing the input vacuum state is not possible. However, the width
of the distribution $\sigma=\sqrt{(\Delta n_1)^2}$ decreases $\sim
e^{-r}$ with increasing squeezing parameter $r$. Particularly, for a
reasonably large squeezing  $r\sim 1$ noise may be reduced by more
than $50\%$.

\begin{figure}
\unitlength1cm
\begin{picture}(18,7)(0,0)
\put(0,0)     {\includegraphics[width=10.cm]{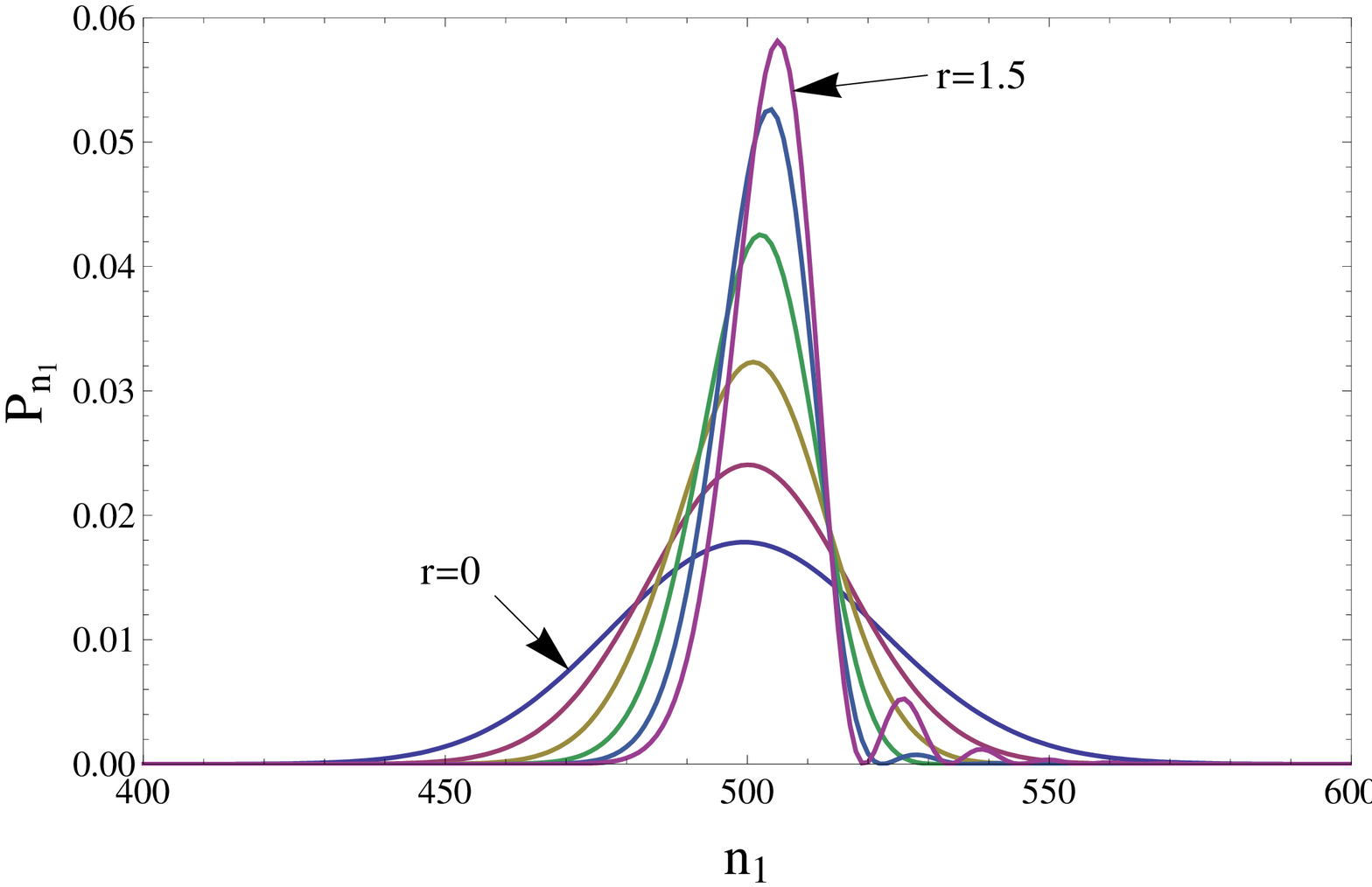}}
\put(10.1,3.5){\includegraphics[width=4.cm] {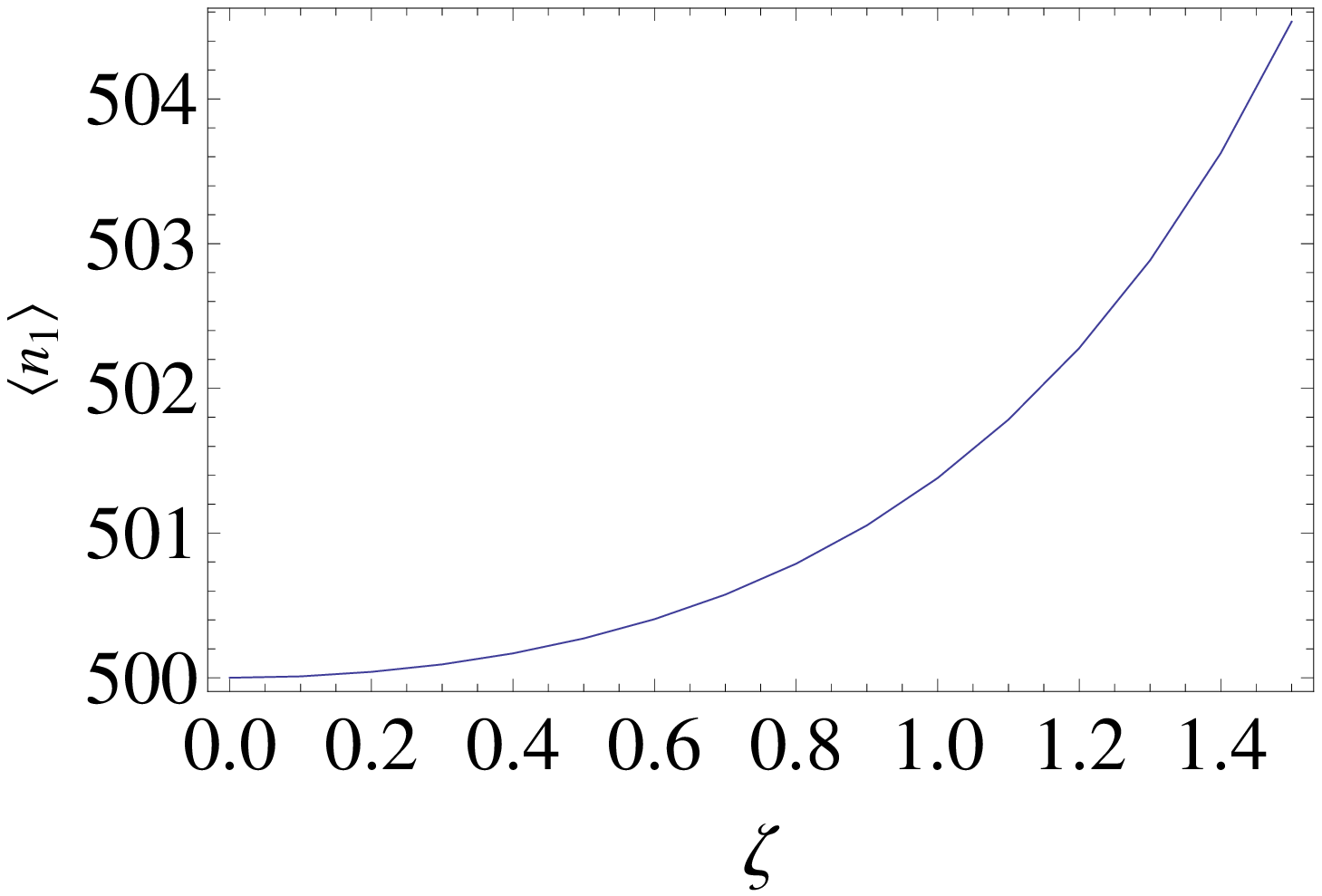}}
\put(10,.5)   {\includegraphics[width=4.cm] {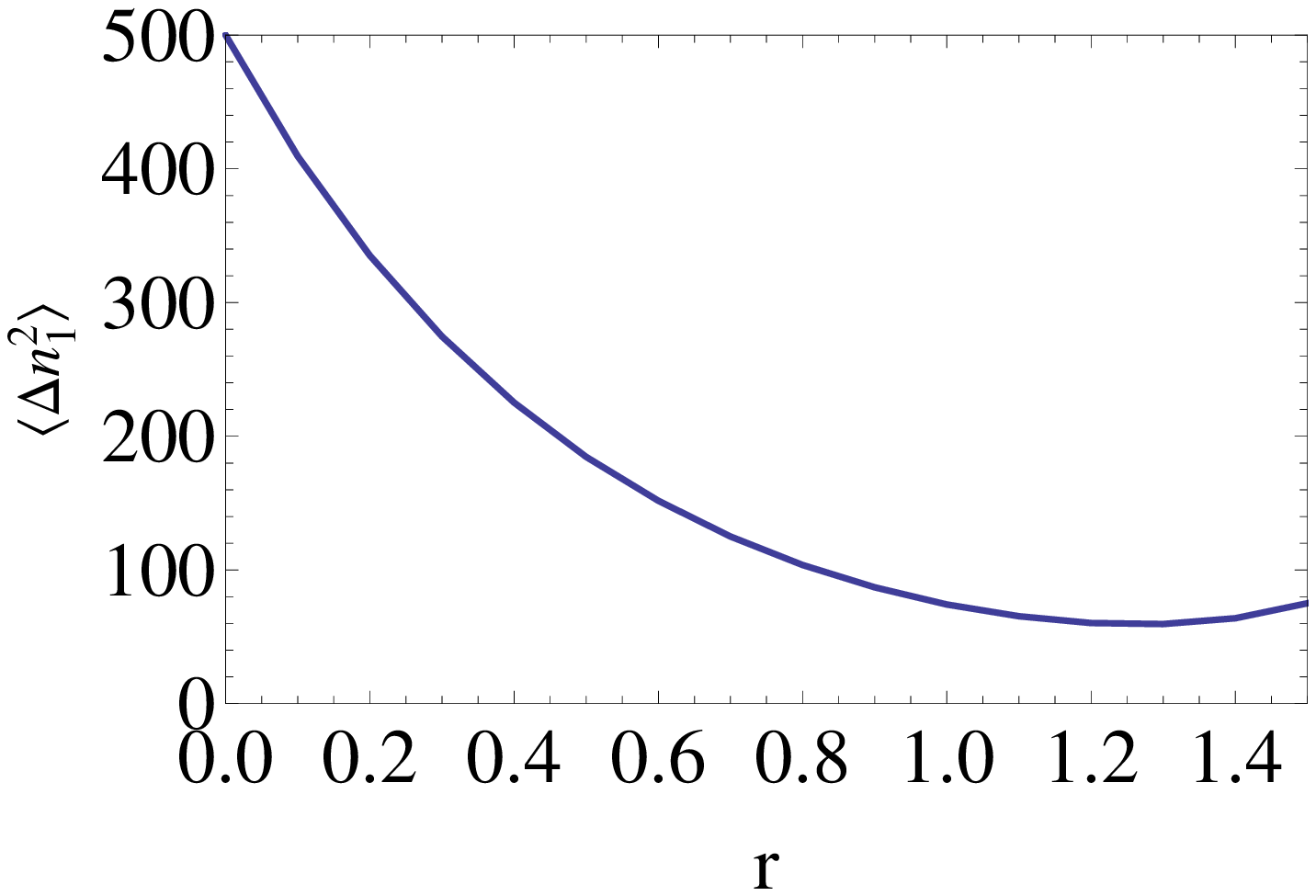}}
\end{picture}
\caption{(Color online) Photon number probability distribution for $|\delta\alpha|^2=500$, $\theta=2\phi$, and different
values of the squeezing parameter $r=0,~0.3,~0.6,~0.9,~1.2,~1.5$ in the dark port.
On the right hand side of the plot we show the mean and the variance squared of these distributions.
}\label{fig:phot_distr500}
\end{figure}

In Fig.~\ref{fig:phot_distr500} we show for the case
$\theta-2\phi=0$ and $|\delta\alpha|^2=500$ the photon number
distribution of the output state calculated from
Eq.~(\ref{eq:darkdist}). Additionally we determine the mean and the
variance of these distributions from Eq.~(\ref{eq:meanvari_case}).
Notice, that for large squeezing parameters the distributions show
characteristic oscillations~\cite{scully:1}.

\section{Summary and conclusions}

In this paper we studied the entanglement effects on the photon
distributions in the output of the interferometer for the `dark
port' case, when the beam splitter is oriented close to $90^{\circ}$
to an incoming coherent state and a squeezed vacuum state is
injected into the usually unused second input port.

Our results for the `dark port' case show that squeezing  does not
influence the mean of the distribution tangibly, thus there is no
amplification of the signal. This result contrasts with the findings
of Ref.~\cite{barak:1}. Signal amplification can only be achieved by
increasing the intensity of the input coherent state, that is by
increasing of  $|\alpha|$. Squeezing allows to decrease the noise:
for a squeezing factor $r\sim 1$ the reduction of noise is more then
$50\%$.

Furthermore, our analysis shows that the mean of the distribution
does not depend on the phases $\theta$ (squeezing parameter phase)
or $\phi$ (coherence parameter phase), but the variance of the
distribution depends on the phase relation $\theta-2\phi$. The most
appropriate choice in order to achieve minimum possible counting
noise is $\theta-2\phi=0$.

\vskip 2cm

{\bf Acknowledgements}

We would like to thank Yacob Ben-Aryeh for a useful correspondence.
M. W. acknowledges Roman Schnabel for a useful correspondence and
Robert Wynands for useful discussions on experimental issues. V.G.V.
thanks Physikalisch-Technische Bundesanstalt for support of an
internship, while this work was performed.

\end{document}